\documentclass{article}
\usepackage{spconf,amsmath,graphicx,hyperref}

\usepackage{leadsheets}     % Lead sheets and songs
\usepackage{amsmath}
\usepackage{amssymb}
\usepackage[T1]{fontenc}    % use 8-bit T1 fonts
\usepackage{hyperref}       % hyperlinks
\usepackage{booktabs}       % professional-quality tables
\usepackage{amsfonts}       % blackboard math symbols
\usepackage{pifont}
\usepackage{nicefrac}       % compact symbols for 1/2, etc.
\usepackage{microtype}      % microtypography
\usepackage[dvipsnames]{xcolor}         % colors
\usepackage{fontawesome5}
\usepackage{mwe}
\usepackage{bm}
\usepackage{enumitem}
\usepackage{cite}
\usepackage[usestackEOL]{stackengine}
\usepackage{multicol}
\usepackage{multirow}
\PassOptionsToPackage{hyphens}{url}
\hypersetup{                    % ...like so
  colorlinks=true,
  linkcolor={RedViolet},
  citecolor={OliveGreen},
  urlcolor={BlueViolet}
}

\newcommand{\cmark}{{\color{ForestGreen} \ding{51}}}
\newcommand{\xmark}{{\color{Maroon}\ding{55}}}

\newcommand{\fref}[1]{Fig.~\ref{#1}}

\definecolor{LightGray}{gray}{0.6}

\title{Stemphonic: All-at-once Flexible Multi-stem Music Generation}
\name{Shih-Lun Wu\thanks{* Work done while an intern at Adobe Research.}$^{\,\text{\sharp\:\flat*}}$ \quad Ge Zhu$^{\,\text{\flat}}$ \quad Juan-Pablo Caceres$^{\,\text{\flat}}$ \quad Cheng-Zhi Anna Huang$^{\,\text{\sharp}}$ \quad Nicholas J.~Bryan$^{\,\text{\flat}}$}
\address{$^{\text{\sharp}}\,$ MIT CSAIL \quad \quad $^{\text{\flat}}\,$ Adobe Research
}
\begin{document}
\ninept
\maketitle

%%%%%%%%%%%%%%%%%%%%%%%%%%%%%%%%%%%%%%%%%%%%%%%%%%%%%%%%%%%%%%%%%%%%%%%%%%%%%%%%
%%%%%%%%%%%%%%%%%%%%%%%%%%%%%%%%%%%%%%%%%%%%%%%%%%%%%%%%%%%%%%%%%%%%%%%%%%%%%%%%
%%%%%%%%%%%%%%%%%%%%%%%%%%%%%%%%%%%%%%%%%%%%%%%%%%%%%%%%%%%%%%%%%%%%%%%%%%%%%%%%
\begin{abstract}
Music stem generation, the task of producing musically-synchronized and isolated instrument audio clips, offers the potential of greater user control and better alignment with musician workflows compared to conventional text-to-music models.
Existing stem generation approaches, however, either rely on fixed architectures that output a predefined set of stems in parallel, or generate only one stem at a time, resulting in slow sequential inference despite flexibility in stem combination. 
We propose \textsc{Stemphonic},
a diffusion-/flow-based framework that overcomes this trade-off and generates a variable set of synchronized stems in one inference pass.
During training, we treat each stem as a batch element, group synchronized stems in a batch, and apply a shared noise latent to each group. 
At inference-time, we use a shared initial noise latent and stem-specific text inputs to generate synchronized multi-stem outputs in one pass.
We further expand our approach to enable one-pass conditional multi-stem generation and stem-wise activity controls to empower users to iteratively generate and orchestrate the temporal layering of a mix.
We benchmark our results on multiple open-source stem evaluation sets and show that \textsc{Stemphonic} produces 
higher-quality outputs while accelerating the full mix generation process by 25--50\%. 
Demos at: {\footnotesize \url{https://stemphonic-demo.vercel.app}}.

% process to compose a full mix.
\end{abstract}

%%%%%%%%%%%%%%%%%%%%%%%%%%%%%%%%%%%%%%%%%%%%%%%%%%%%%%%%%%%%%%%%%%%%%%%%%%%%%%%%
%%%%%%%%%%%%%%%%%%%%%%%%%%%%%%%%%%%%%%%%%%%%%%%%%%%%%%%%%%%%%%%%%%%%%%%%%%%%%%%%
%%%%%%%%%%%%%%%%%%%%%%%%%%%%%%%%%%%%%%%%%%%%%%%%%%%%%%%%%%%%%%%%%%%%%%%%%%%%%%%%
\begin{keywords}music audio generation, stem generation, conditional stem generation, variable stem combinations, diffusion, flow.
\end{keywords}

%%%%%%%%%%%%%%%%%%%%%%%%%%%%%%%%%%%%%%%%%%%%%%%%%%%%%%%%%%%%%%%%%%%%%%%%%%%%%%%%
%%%%%%%%%%%%%%%%%%%%%%%%%%%%%%%%%%%%%%%%%%%%%%%%%%%%%%%%%%%%%%%%%%%%%%%%%%%%%%%%
%%%%%%%%%%%%%%%%%%%%%%%%%%%%%%%%%%%%%%%%%%%%%%%%%%%%%%%%%%%%%%%%%%%%%%%%%%%%%%%%
\section{Introduction}
\label{sec:intro}

% P1: Convention music gen models are limiting
Text-to-audio music generation models are now able to produce realistic sounding music from simple text inputs~\cite{agostinelli2023musiclm,copet2023simple,evans2024fast,novack2025presto,team2025live}.
They lower the barrier for music creation, enabling anyone to explore and express their creativity, but typically generate \textit{fully-mixed} multi-instrument outputs that are difficult to edit and cannot easily be reused as components in new compositions~\cite{donahue2025hookpad,kim2025amuse}.
To empower creators beyond text prompting, numerous control and editing methods have been proposed, including fine-grained temporal controls~\cite{wu2024music,tal2024joint,tsai2025musecontrollite}, music inpainting~\cite{novack2024ditto,tsai2024audio,tsai2025musecontrollite}, 
% length extension, loop generation, 
as well as music stem generation~\cite{ mariani2024multi,yao2025jen,karchkhadze2025simultaneous,rouard2025musicgen}.
A music \emph{stem} is a recording of one or more instruments that collectively serve as a distinct layer in a mix, e.g., \textit{drums} for rhythmic foundation, and \textit{basses} for low-pitch progressions.
Generating stems enables creators to edit each stem separately and experiment with different mixing/mastering techniques, enhancing their creative control~\cite{malandro2023composer,malandro2024composer}. 

% P2: Past music stem gen methods 
Existing stem generation methods can largely be classified into 
(\textit{1})~models with a \textit{parallelized} architecture,
and 
(\textit{2})~individual-stem models that require \textit{sequential} generation of stems.
% designs.
Parallelized models~\cite{mariani2024multi,yao2025jen,karchkhadze2025simultaneous,rouard2025musicgen} generate \textit{multiple} coherent stems in a \textit{single pass}, but handle limited, coarse-grained stem types (e.g., only basses, drums, vocals, others) that must be fixed in advance and built into the architecture.
Individual-stem models~\cite{han2024instructme,parker2024stemgen,nistal2024diff,chae2025mge,strano2025stage}, on the other hand, allow flexible open-vocabulary stem generations through text prompting or other conditioning mechanisms, and can often condition on existing audio to iteratively generate new accompanying stems and create a mix with an arbitrary number of stems.
% enabling an ability to iteratively create a mix via adding a variable number of stems together. 
% Sequential models, however, generate stems \textit{one at a time},
% leading to slower inference time.
Yet, these models generate stems \textit{one at a time},
leading to slower full inference processes.

\begin{figure}[t!]
    \centering
    \includegraphics[width=0.96\linewidth,trim={0 0 0 0},clip]{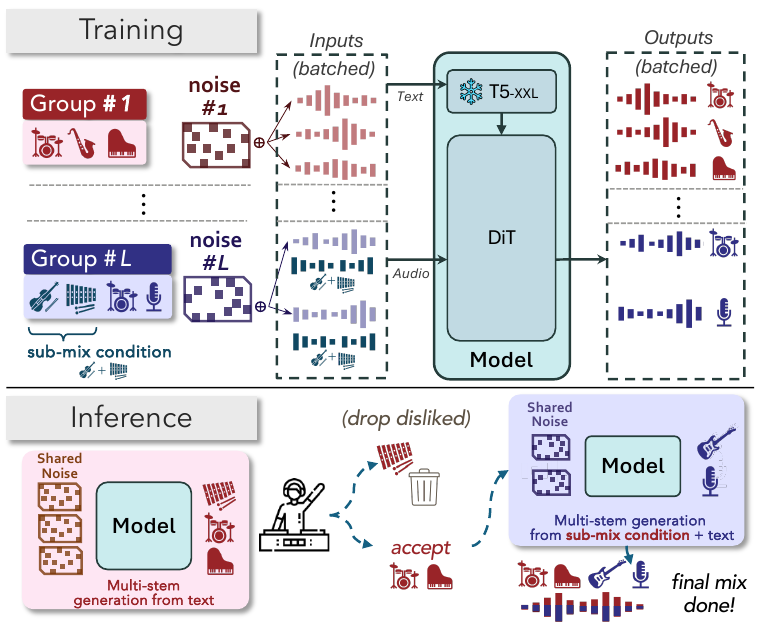}
    \vspace{-2mm}
    \caption{Our \textsc{Stemphonic} framework for flexible multi-stem music generation. 
    (Top) At training, each group of synchronized stems share the same noise latent.
    (Bottom) At inference, we use a shared initial noise to generate variable multi-stem outputs in one pass. We also enable conditional stem generation and stem-wise activity controls.}
    \vspace{-4mm}
    \label{fig:headline}
\end{figure}

% P3: Our approach
To unify the strengths of both paradigms
and alleviate their drawbacks,
we propose \textsc{Stemphonic},
a latent diffusion/flow-based~\cite{liu2023flow,esser2024scaling} framework capable of generating a variable set of musically-synchronized stems in one inference pass,
as shown in~\fref{fig:headline}. 
% To do so,
We introduce two techniques applied during training. 
First, we treat each stem as a batch element and group musically-synchronized stems in a batch (Sec.~\ref{subsec:data-group}).
Then, we assign a single shared noise latent to each group (Sec.~\ref{subsec:noise-share}). 
At inference, we use a shared initial noise and different stem-specific text inputs to generate variably many synchronized stem outputs in one pass.
We further expand our approach to conditional multi-stem generation, and add stem-wise activity controls (Sec.~\ref{subsec:actcurve}) that enable creators to iteratively generate and orchestrate the temporal layering of a mix. 
In our experiments, we ablate our stem grouping and noise sharing core techniques,
and verify the conditional generation and stem-wise activity control capabilities.
We find \textsc{Stemphonic} capable of generating higher-quality multi-stem mixes, while accelerating the full inference process by 25--50\%, compared to the existing individual-stem iterative workflow.
% We find our approach is effective at one-pass inference of multiple musically-synchronized outputs, conditional stem generation, and activity control. 

% \item A method to enable conditional stem generation for iterative, multi-pass, multi-track stem generation in one inference pass.

% In total, our contributions are:
% \noindent
Our contributions can be summarized as follows:
\begin{itemize}[leftmargin=*,topsep=2pt,itemsep=1pt]
    \item A latent diffusion/flow-based framework to generate a variable number of synchronized stems efficiently in one inference pass, either from-scratch, or conditioned on existing audio.
    \item A method for precise activity control of individual stem outputs.
    \item Quantitative evaluation and qualitative demo examples showing high-quality outputs, and flexible composer-like workflows.
\end{itemize}
Overall, we combine the speed of parallelized models with the open-vocabulary freedom of individual-stem models to offer an efficient and highly-controllable stem generation framework.

\section{Background}\label{sec:bg}
% Latent diffusion/flow models
We build upon latent diffusion~\cite{rombach2022high} and rectified flow (RF)~\cite{liu2023flow} generative models with a Transformer backbone~\cite{peebles2023scalable}, referred to as diffusion Transformer (DiT) for simplicity. 
% Our DiT model is initialized from a pre-trained base model trained 
% Our DiT model is pretrained on general music audios (i.e., not specifically stems), and then finetuned on stems
We initialize our DiT with weights from a base model pretrained on general music mixes,
and then finetune it on isolated stems 
% according to the objective below:
with the following RF objective:
\begin{equation}    \label{eq:plain-rf}
   \min_\theta \,\mathop{\mathbb{E}}_{\substack{(\bm{x}_k, \mathcal{C}_k) \: \sim \: \text{data} \\ 
\bm{\epsilon}\:\sim\:\mathcal{N}(0,\, \mathbf{I}_{\mathrm{T}D \times \mathrm{T}D }) \\ 
   t \: \sim \: \text{LogitNormal$_{\mu, \sigma}$} \in (0, 1)}} {\Big \lVert} \bm{v}_\theta \big( \bm{x}_{k}(t), t, \mathcal{C}_k \big) - \big(\bm{x}_k - \bm{\epsilon}
   \big) {\Big\rVert}^2_2 \, ,
\end{equation}
where $\bm{x}_k \in \mathbb{R}^{\mathrm{T} \times D}$ is the variational autoencoder (VAE) audio latents being modeled,
$k$ is the stem index,  
$\mathrm{T}$ is the number of VAE latent frames (i.e., the \textit{time} dimension),
$D$ is the frame-wise dimensionality,
$\theta$ is the set of trainable DiT parameters,
$\mathcal{C}_k$ is any arbitrary conditions associated with stem $\bm{x}_k$ that enable user control (e.g., stem type, audio mix of existing stems, text description, and tempo),
$\bm{\epsilon} \in \mathbb{R}^{\mathrm{T} \times D}$ is the gaussian initial noise,
$t \in (0, 1)$ is the sampled noise level (i.e., diffusion/flow timestep),
$\bm{x}_{k}(t) := (1-t)\bm{x}_k + t\bm{\epsilon}$ is the noised VAE latents the model receives,
and $\bm{v}_{\theta}(\cdot)$ is the velocity predicted by the model to \textit{denoise} the latents.
During inference, we solve the associated probability flow ODE,
$d\bm{x}_{k}(t)=\bm{v}_\theta\big(\bm{x}_{k}(t), \mathcal{C}_k, t\big)dt$,
which deterministically transports the initial gaussian noise to realistic data via Euler discretization~\cite{song2020denoising}.

\section{Method}\label{sec:method}
We propose a novel framework capable of generating a variable number/set of musically-synchronized stems in one pass.
We achieve this with two main techniques to intervene in training batch construction,
i.e., \textit{stem grouping} and \textit{noise sharing},
which instill an inductive bias
for inter-stem cohesion and synchronization directly into the model.

% We propose a novel method for
% generating multiple musically-synchronized stems, with variable stem count,
% in one inference pass.
% To advance from generating independent stems,
% as characterized in Eqn.~\eqref{eq:plain-rf},
% to a cohesive, multi-stem mix, our goal is to introduce a mechanism that models their correlation and eventually generate multitrack musically synchronized outputs.
% We achieve this through two main techniques that modify how data and noise are presented to the model in training batches
% so as to introduce a mechanism that models 
% to bring about an inductive bias promoting
% inter-stem correlation and cohesion.
% and eventually generate multitrack musically synchronized outputs. 

% Our task is super clear now,
% how to massage the stem-wise independent
% objective laid out in Eqn.~\eqref{eq:plain-rf}
% to enable simultaneous generation of correlated and aligned stems!

%%%%%%%%%%%%%%%%%%%%%%%%%%%%%%%%%%%%%%%%%%%%%%%%%%%%%%%%%%%%%%%%%%%%%%%%%%%%%%%%
%%%%%%%%%%%%%%%%%%%%%%%%%%%%%%%%%%%%%%%%%%%%%%%%%%%%%%%%%%%%%%%%%%%%%%%%%%%%%%%%
\subsection{Stem data \& grouping}\label{subsec:data-group}
% \textbf{Gist:
% Do something about the} 
% ``$(\bm{x}^{(k)}, C^{(k)}) \, \sim \text{data}$'' \textbf{process.}

% Having aligned stems as batch elements is the first necessary step for us to generate a variable/arbitrary set of stems in one pass.
We finetune the DiT on a dataset of isolated stem audio clips.
Stems from the same composition can be linearly combined to form a cohesive musical \textit{mix} $\bm{w}_{\text{mix}} := \sum_k (g_k \cdot\bm{w}_k)$, where $g_k \in \mathbb{R}$ is the user-adjustable \textit{gain} for each individual stem,
and $\bm{w}_{\text{mix}}, \bm{w}_k$ are raw audio waveforms of the mix and each stem.
We further define \emph{sub-mix} as the mixture of a subset of one or more stems in a full mix.
We process individual stereo 44.1kHz waveforms of stems, mixes, and sub-mixes 
% as typical mixtures 
using the same VAE audio-to-latent encoder and latent-to-audio decoder~\cite{evans2024fast,yao2025jen,karchkhadze2025simultaneous, casebeer2025},
obtaining the VAE latents, i.e., $\bm{x}_k$'s,  for stem finetuning.
% Given a dataset of stem data,
We develop a \textit{stem grouping} mechanism ensuring that
% to finetune our model 
% with 
musically-synchronized stems 
% stem mixes or sub-mixes that naturally 
appear together in a single training batch. 
This is in contrast to the standard batch construction mechanism where individual stems are independently sampled~\cite{parker2024stemgen,nistal2024diff,chae2025mge}.

% That is, instead of creating batches by sampling target stems independently as in~\cite{parker2024stemgen,nistal2024diff,chae2025mge},
% we ensure our batch contains groups of aligned/compatible stems. 

We let $\mathcal{M}$ be our training dataset of multi-stem compositions (or mixes),
in which each mix is denoted by its constituent stems' VAE latents $\mathcal{X}^{(m)} = \{\bm{x}^{(m)}_k\}^{K^{(m)}}_{k=1}$,
where $m \in \{1, \dots, |\mathcal{M}|\}$ indexes the mix and $K^{(m)}$ is the $m^{\text{th}}$ mix's number of constituent stems.
% \slwu{Perhaps the sentence above should be in Sec.~\ref{sec:bg}.}
We also let $\mathcal{B}$ be a training batch (of stems) to be sampled, with a fixed batch size $B$.
To construct $\mathcal{B}$, we run a loop where in
each iteration, indexed by $l \in \mathbb{N}$, we sample a mix $m(l) \sim\{1, \dots,|\mathcal{M}|\}$, and then sample a subset of the mix's constituent stems $\tilde{\mathcal{X}}^{(l)} \subseteq \mathcal{X}^{(m(l))}$ to obtain $|\tilde{\mathcal{X}}^{(l)}|$ target stems as batch elements. 
% that are separate in terms of the model's computation graph.
The loop runs until the number of target stems reaches batch size $B$, and hence the number of groups per batch, denoted by $L$, is dependent on the sampled groups.
We sample subsets from a mix, rather than always including the entire mix, to combinatorially increase data variability.

% For each batch, there is by definition a one-to-one mapping from the sampled groups to the dataset of mixes ($\{1, \dots, L\} \rightarrow \{1, \dots, |\mathcal{M}|\}$)

To support iterative creation workflows,
we also train our model to generate new stems conditioned on an existing stem or sub-mix, similar to~\cite{nistal2024diff}.
% \slwu{We can also write in math notations below but I feel that will add a lot more symbols and perhaps not required for understanding.}
Operationally, we randomly select half of the $L$ groups in a batch for this \textit{conditional generation} task. 
For those selected groups,
we sample a subset from the `left out' stems of the corresponding mix, i.e., $\mathcal{X}^{(m(l))} \setminus \tilde{\mathcal{X}}^{(l)}$, sum their waveforms together, and then obtain the VAE latents of the sub-mix as the condition, i.e., $\bm{x}^{(l)}_{\text{cond}} \in \mathbb{R}^{\mathrm{T} \times D}$.
Note that this is effectively treating the sub-mix as part of the conditions $\mathcal{C}_k$ (cf.~Eqn.~\eqref{eq:plain-rf}) that is shared among each output stem in a group $\bm{x}^{(l)}_k \in \tilde{\mathcal{X}}^{(l)}$.
This condition is channel-wise concatenated
% we concatenate $\bm{x}^{(l)}_{\text{cond}}$ channel-wise 
to the noised target stem latents  $\bm{x}^{(l)}_{k}(t)$ as the input to our DiT model.
% If we are not using the condition, we impute zeros for these channels.
For groups that are not selected 
(i.e., those training for the \textit{from-scratch generation} task),
we impute $\bm{x}^{(l)}_{\text{cond}}$ with zeros.
% \slwu{passed sec 2--here.}

% We use equal weighting on mixes, i.e., each mix contributes the same number of stems (sampled with replacement) in a batch.
% To get more diverse data, we do subset sampling, rather than including all stems every time.

%%%%%%%%%%%%%%%%%%%%%%%%%%%%%%%%%%%%%%%%%%%%%%%%%%%%%%%%%%%%%%%%%%%%%%%%%%%%%%%%
%%%%%%%%%%%%%%%%%%%%%%%%%%%%%%%%%%%%%%%%%%%%%%%%%%%%%%%%%%%%%%%%%%%%%%%%%%%%%%%%
\vspace{-1mm}
\subsection{Diffusion/flow noise sharing}\label{subsec:noise-share}
\vspace{-1mm}
% \textbf{Gist:
% Do something about the} 
% ``$\bm{\epsilon} \,\sim\,\mathcal{N}(0,\, \mathbf{I}_{\mathrm{T} \times D})$'' \textbf{process.}

% One issue of the data grouping introduced above 
Even with stem grouping,
a remaining challenge is that
multiple stem groups, each corresponding to a distinct mix, are present in the same batch.
Since individual stems are each a batch element,
in the model's computation graph,
every stem is separate regardless of the groupings.
The only shared information among stems in the same group comes from portions of $\mathcal{C}_k$, which include parts of the text prompt (more details in Sec.~\ref{subsec:text-ctrl}), and the conditioning sub-mix $\bm{x}^{(l)}_{\text{cond}}$ in conditional generation cases. 
Therefore, additional signals are required to more strongly inform 
the grouping information for the model to learn the inter-stem alignment and orchestration well.
% \slwu{edited sec 2 to here.}

To achieve this, we intervene in the noise sampling process. 
The initial noise $\bm{\epsilon}$ is a high-dimensional variable that provides the source of diversity for generation after conditioning on $\mathcal{C}_k$.
Conventionally, $\bm{\epsilon}$ is sampled independently for batch elements assuming
% batch elements 
they
are \textit{i.i.d.}
However, since our batches now contain groups of correlated stems,
% and given the high degree of freedom 
we can leverage the noise latent's high dimensionality
($\mathbb{R}^{\mathrm{T} \times D}$) as a powerful signal to indicate groupings.
% we reason that the noise is a suitable component to intervene.
Specifically, 
% we sampled  pairing a single noise 
for each group $l \in \{1, \dots,L
\}$,
we sample a shared initial noise $\bm{\epsilon}^{(l)}$ to be applied to all stems in the group,
i.e., the noise is paired to stems as:
\begin{equation}
    (\bm{\epsilon}^{(l)}, \bm{x}_k^{(l)}) \;\; \forall \, \bm{x}_k^{(l)} \in \tilde{\mathcal{X}}^{(l)}; \quad \bm{\epsilon}^{(l)} \sim \mathcal{N}(0,\, \mathbf{I}_{\mathrm{T}D \times \mathrm{T}D }) \, .
\end{equation}
By doing so, we ensure that musically-synchronized stems in a group receive a shared noise,
while different groups (which satisfy \textit{i.i.d.}) still receive independent noises,
strongly enforcing the grouping property.
% enforcing strong signals of grouping on the model.
At inference, we assign one random initial noise shared across all batch elements to generate musically-synchronized stems.
Although implemented here with the RF objective, our noise sharing technique is generally applicable to any diffusion- or flow-based model that utilizes a high-dimensional noise latent.

% Diffusion initial noise is a good choice since the space is as large as the VAE latents ($\mathbb{R}^{\mathrm{T} \times D}$), and hence will give a strong/pronounced hint to the model.

% Write out an equation characterizing the assignment of noises.

% \textbf{Algorithm?}---maybe having some pseudocode summarizing the operations in the 2 subsections above would be helpful

%%%%%%%%%%%%%%%%%%%%%%%%%%%%%%%%%%%%%%%%%%%%%%%%%%%%%%%%%%%%%%%%%%%%%%%%%%%%%%%%
%%%%%%%%%%%%%%%%%%%%%%%%%%%%%%%%%%%%%%%%%%%%%%%%%%%%%%%%%%%%%%%%%%%%%%%%%%%%%%%%
% \vspace{-}
\vspace{-1mm}
\subsection{Stem-wise activity control}\label{subsec:actcurve}
\vspace{-1mm}
% \textbf{Gist: add something useful and novel to} $C^{(k)}$.
In addition to generating well-synchronized stems,
to grant users more compositional agency, especially when generating a large number of stems (e.g., $>5$), we devise a mechanism for stem-wise activity control.
This empowers users to precisely layer the final mix by specifying the temporal activity of each stem.

Operationally,
we run simple loudness-based silence detection on the raw waveform of each stem, $\bm{w}_k$.
Based on a preset cutoff loudness (e.g., $-60$ dB),
we obtain a list of silence segments.
We then convert/quantize it into a binary, single-channel sequence denoted by
$\bm{a}_k \in \{0, 1\}^{\mathrm{T} \times 1}$,
where $1$ indicates active and $0$ silent,
and add it to the conditions $\mathcal{C}_k$.
(Note the time dimension $\mathrm{T}$ is consistent with that of target stem latents $\bm{x}_k$.)
We learn a small (16-dim.) embedding for activity indicators $1$ and $0$,
and concatenate the sequence of embeddings channel-wise to the noised target $\bm{x}_{k}(t)$
before entering the DiT,
 mirroring the sub-mix conditioning mechanism (cf.~Sec.~\ref{subsec:data-group}).

% Method-wise: Energy-based silence segment detection,
% and then transformed/quantized to binary sequence aligned to VAE latents,
% and then channel-concat to (noised) VAE latents in a similar fashion to the conditioning audio in cond gen.

%%%%%%%%%%%%%%%%%%%%%%%%%%%%%%%%%%%%%%%%%%%%%%%%%%%%%%%%%%%%%%%%%%%%%%%%%%%%%%%%
%%%%%%%%%%%%%%%%%%%%%%%%%%%%%%%%%%%%%%%%%%%%%%%%%%%%%%%%%%%%%%%%%%%%%%%%%%%%%%%%
\subsection{Text conditioning}\label{subsec:text-ctrl}
% \textbf{Gist: ensure the text part in} $C^{(k)}$ \textbf{make sense}.
Our model supports free text control as in conventional text-to-music models~\cite{copet2023simple,evans2024fast,team2025live}.
Compared to prior art on text-conditioned individual stem generation~\cite{han2024instructme,nistal2024diff,chae2025mge},
or fixed-combination parallelized stem generation~\cite{yao2025jen,rouard2025musicgen}, our method requires combining both \textit{global} (i.e., shared across stems in a group) and \textit{stem-wise} 
descriptions in one text instruction.
We formalize this by placing the {\color{Plum}{\textit{stem-wise}}} part first, followed by the {\color{Bittersweet}{\textit{global}}} part.
An example prompt is:
``\textit{Generate the stem: {\color{Plum} [GUITARS]} {\color{Bittersweet}{(given context stems: [DRUMS, BASSES])}} for some music described as: {\color{Bittersweet}{Relaxing country rock \dots}}}''.
The `\textit{given context stems}' clause is only used for conditional generation (i.e., not for from-scratch cases) and is randomly dropped out at training.

% \slwu{can move this section to under Sec.~\ref{subsec:data-group} if low on space.}

% \slwu{first pass until here.}

% And we do this by specifying the target stem first, and the global style text description.

% \slwu{This can be combined into Sec.~\ref{subsec:impl-details} if low on space.}

%%%%%%%%%%%%%%%%%%%%%%%%%%%%%%%%%%%%%%%%%%%%%%%%%%%%%%%%%%%%%%%%%%%%%%%%%%%%%%%%
%%%%%%%%%%%%%%%%%%%%%%%%%%%%%%%%%%%%%%%%%%%%%%%%%%%%%%%%%%%%%%%%%%%%%%%%%%%%%%%%
%%%%%%%%%%%%%%%%%%%%%%%%%%%%%%%%%%%%%%%%%%%%%%%%%%%%%%%%%%%%%%%%%%%%%%%%%%%%%%%%
\section{Experiments}\label{sec:expr}

%%%%%%%%%%%%%%%%%%%%%%%%%%%%%%%%%%%%%%%%%%%%%%%%%%%%%%%%%%%%%%%%%%%%%%%%%%%%%%%%
%%%%%%%%%%%%%%%%%%%%%%%%%%%%%%%%%%%%%%%%%%%%%%%%%%%%%%%%%%%%%%%%%%%%%%%%%%%%%%%%
\subsection{Implementation details}\label{subsec:impl-details}
Our model architecture is similar to that of Stable Audio Open~\cite{evans2025stable}---A VAE~\cite{casebeer2025} compresses stereo 44.1 kHz waveforms into latents (with $D = 64$ dimensions) at a 12 Hz frame rate.
A billion parameter-sized diffusion transformer (DiT)~\cite{peebles2023scalable} models the VAE latents,
and text conditions enter the DiT as T5-XXL~\cite{raffel2020exploring} embeddings via cross attention.
Besides the conditions introduced in Sec.~\ref{sec:method}, the model is also conditioned on tempo using bpm values.
All conditions are independently dropped out 1/3 of the time to enable classifier-free guidance (CFG)~\cite{ho2022classifier}.
We pretrain on full audio mixes before finetuning on stems with our proposed techniques.
The model is trained on 32-second segments (equates to \# of frames $\mathrm{T} = 394$), with a per-GPU batch size of 1024 seconds,
and an effective batch size (per gradient step) of 16K seconds.
We use AdamW optimizer~\cite{loshchilov2017decoupled} with a constant $10^{-4}$ lr.
Stem generation training converges at 30K gradient steps,
which takes 3 days on 8$\times$ A100 (80G) GPUs.
At inference,
we use a first-order Euler sampler
with 32 steps.
We apply CFG to all conditions only at steps 3$\sim$28~\cite{kynkaanniemi2024applying}, with the CFG scale set to 3.0.

% some description on model architecture (VAE \slwu{just say it's similar to Stable Audio~\cite{evans2024fast}?}, DiT~\cite{peebles2023scalable}, cross attend to text, BPM conditioning, condition dropouts)
% training process (learning rate, batch size, GPU budget, number of steps)
% and inference setup (Euler sampler, limited CFG)
% T5 text reprs~\cite{raffel2020exploring}.

% Mention that, following prior work~\cite{mariani2024multi, karchkhadze2025simultaneous}, we add separation as an additional task. Even though that's not our goal, we find it improving the stem generations

% \slwu{need guidance on how much we can write?}

%%%%%%%%%%%%%%%%%%%%%%%%%%%%%%%%%%%%%%%%%%%%%%%%%%%%%%%%%%%%%%%%%%%%%%%%%%%%%%%%
%%%%%%%%%%%%%%%%%%%%%%%%%%%%%%%%%%%%%%%%%%%%%%%%%%%%%%%%%%%%%%%%%%%%%%%%%%%%%%%%
\subsection{Datasets}\label{subsec:datasets}
% For training, we use a licensed dataset of stems containing $\sim$400 hours of mixes and their corresponding stems ($\approx6$/mix).
We pretrain on 20K hours of licensed music mixes, and then finetune on licensed stems corresponding to $\sim$400 hours of mixes (with ${\approx}6$ stems/mix).
% Each piece is 2.5 minutes long on average.
Both datasets contain mix-level text descriptions and tempo (bpm) metadata,
with an average track duration of 2.5 minutes.
While our stem dataset comprises more than 50 stem types, the following 11 types constitute the vast majority: drums, basses, percussion, synths, keys, guitars, strings, SFX, vocals, synth-vocals, and winds. 
For evaluation,
following recent literature~\cite{chae2025mge}, we use open-source stem-separated datasets: {MoiseDB}~\cite{pereira2023moisesdb} and MusDB~\cite{musdb18}, both consisting of around 10 hours of mixes.
To construct evaluation examples,
we crop the mixes into 32-second segments,
and include all stems with $\geq$50\% activity in the segment.
We leverage Qwen2.5-Omni~\cite{xu2025qwen2}, a multimodal large language model with music analysis abilities,
to obtain the text descriptions,
and an in-house version of Madmom beat tracker~\cite{bock2016madmom} to estimate the tempo (bpm).
We map the stem types in MoiseDB and MusDB to our top-11 stem types above largely with rules; for those originally tagged `other*', we resort to Qwen2.5-Omni labeling and then manual checking/correction.
The top ($>$100 occurrences) stem types in the  MoisesDB dataset include: drums, basses, guitars, vocals, piano, keys, percussion, and string.
For MusDB, the top stem types are drums, basses, vocals, guitars, keys, and synths.
% in each evaluation dataset are:
% \begin{itemize}[leftmargin=*]
%     \item \textbf{MoisesDB}: {\footnotesize \texttt{[DRUMS, BASSES, GUITARS, VOCALS, PIANOS, \\ KEYS, PERCUSSION, STRINGS]}}
%     \item \textbf{MusDB}: {\footnotesize \texttt{[DRUMS, BASSES, VOCALS, GUITARS, KEYS, SYNTHS]}}
% \end{itemize}
After segmentation,
we get around 1.5K mixes (i.e., grouped stems) for MoisesDB, and about 1K for MusDB.
The mean \# of stems per mix is 4.6 (stdev 1.3) for MoisesDB, and 3.5 (stdev 0.7) for MusDB.
% - training dataset (stem separated, dataset size???, 11 most-frequent stem types)
% \slwu{edited from sec 2 to here.}

% ###########################
% MoisesDB
% ###########################
% DRUMS         1387
% BASSES        1377
% GUITARS       1324
% VOCALS        1188
% PIANOS         618
% KEYS           484
% PERCUSSION     344
% STRINGS        147
% WINDS           70

% ###########################
% MusDB
% ###########################
% DRUMS         941
% BASSES        894
% VOCALS        749
% GUITARS       682
% KEYS          138
% SYNTHS        103
% STRINGS        54
% SFX            19
% WINDS          14
% PERCUSSION      6

% \noindent
% - eval datasets (MusDB~\cite{musdb18} MoisesDB~\cite{pereira2023moisesdb}): the stem mapping process, and getting global text prompts from Qwen~\cite{xu2025qwen2},

% \textbf{Figure?}---
% at least provide a stem type distribution on eval datasets
% (also best if we can do that for training dset)

% \slwu{need guidance on how much we can write?}

%%%%%%%%%%%%%%%%%%%%%%%%%%%%%%%%%%%%%%%%%%%%%%%%%%%%%%%%%%%%%%%%%%%%%%%%%%%%%%%%
%%%%%%%%%%%%%%%%%%%%%%%%%%%%%%%%%%%%%%%%%%%%%%%%%%%%%%%%%%%%%%%%%%%%%%%%%%%%%%%%
\subsection{Baselines \& experimental setup}
Our experiments are structured in three sets.
% We first set aside stem-wise activity controls, and
% evaluate the core capability of generating variably many coherent stems in \textit{one inference pass}.
For our main ablations,
we study our core capability of generating variably many musically-synchronized stems in \textit{one inference pass}. To do so, we apply stem grouping (Sec.~\ref{subsec:data-group}) and noise sharing (Sec.~\ref{subsec:noise-share}) (we dub this setup `\textbf{C}'),
or only stem grouping (`\textbf{B}'), or neither (`\textbf{A}'), resulting in three trained models.
Note that, whether or not noise sharing is applied at training,
we have the option to share the noise at inference among the stems meant to be synchronized,
as a training-free measure to promote inter-stem coherence.
This gives us two inference-time variations: sharing the noise~(dubbed `\textbf{(ii)}') or not~(`\textbf{(i)}').
% Therefore, we have the inference-time variation of whether to share (dubbed `\textbf{(ii)}'), or not share~(`\textbf{(i)}') the noise at inference.
To summarize,
our full setup is \textbf{C-(ii)},
and the most ablated setup, \textbf{A-(i)}, represents the case of naively stretching existing individual-stem models~\cite{han2024instructme,nistal2024diff,chae2025mge}
(with minimal operational modifications) to multi-stem generation.

Next,
we recognize that
with the basic \textbf{A-(i)} setup,
we can still generate variable $K$-stem mixes by taking $K$ inference passes.
Thus, we need to demonstrate the holistic \textit{workflow improvement} our \textbf{C-(ii)} setup brings about.
% both in terms of output quality and speed.
Thus, we compare generating $K$ stems from scratch using
(\textit{1})~\textbf{$K\:$passes} with \textbf{A-(i)},
(\textit{2})~\textbf{$1\:$pass} with \textbf{C-(ii)}, and
(\textit{3})~\textbf{$2\:$passes} with \textbf{C-(ii)},
a middle ground showing our model's from-scratch and conditional generation abilities,\footnote{Half of the stems are generated in the 1st pass, and the remaining half in the 2nd pass, conditioned on the sub-mix of 1st-pass output stems.
If $K$ is odd, one more stem is assigned to 1st pass.}
and the unique flexibility that lets users take however many (or few) passes they see fit.
Here, we evaluate both the generation quality and total inference time here.

As a final experiment, we train our full \textbf{C} model with the stem-wise activity controls~(Sec.~\ref{subsec:actcurve}).
We then evaluate two key aspects: the effectiveness of the control itself, and its impact on multi-stem generation quality in scenarios both with and without the control applied at inference.
% examine their effectiveness, and check if the multi-stem output quality is preserved, whether or not the user chooses to generate with stem-wise activity controls.
We note that during all inference,
no groundtruth stem audios in the evaluation sets are used;
we only use the derived metadata (i.e., \{text prompt, stem types, tempo\}) as inputs.
All inference is done on 1$\times$ A100 (80G) GPU.
The generated stems are mixed in a way that preserves the relative loudness (between stems) decided by the model,
and then globally normalized to --16 dBFS.

% \slwu{First pass until here.}

% Explain ablations (training) \textbf{A, B, C} and (inference) \textbf{(i), (ii)}

% Explain how we get to the 3 tables and why we design them this way.

%%%%%%%%%%%%%%%%%%%%%%%%%%%%%%%%%%%%%%%%%%%%%%%%%%%%%%%%%%%%%%%%%%%%%%%%%%%%%%%%
%%%%%%%%%%%%%%%%%%%%%%%%%%%%%%%%%%%%%%%%%%%%%%%%%%%%%%%%%%%%%%%%%%%%%%%%%%%%%%%%
\subsection{Evaluation metrics}
We evaluate generated audios according to the following key aspects.
\begin{itemize}[leftmargin=*,itemsep=1pt,topsep=1pt]
    \item \textbf{Stem Control}: We compute the Fr\'echet Audio Distance (FAD)~\cite{kilgour2019fr} between reference and generated \textit{stem audios} using VGGish~\cite{hershey2017cnn} feature extractor.
    We treat all audios of each stem type in MoiseDB or MusDB as a distinct reference set,
    and report the macro-average across all stem types.
    We abbreviate this as \textbf{FAD}$_\text{stem}$.
    \item \textbf{Mix Quality}: We also measure the FAD (shorthand \textbf{FAD}$_\text{mix}$), but here we treat groundtruth \textit{mixes} as the reference.
    Note that this also evaluates stem compatibility since if the stems are high-quality but unsynchronized, the mix would sound unlike any real mixes.\footnote{We also investigate the COCOLA metric~\cite{ciranni2025cocola} for stem compatibility, but find it ineffective, likely due to a domain mismatch between its training data (mainly synthesized audios) and ours (largely studio-recorded audios).}
    \item \textbf{Mix Text Control}:
    We compute the pairwise cosine similarity between the audio embeddings of the generated \textit{mix}, and the text embeddings of global text description, both obtained
    from the contrastively trained \textbf{CLAP} model~\cite{wu2023large}, using the \texttt{music} checkpoint.
\end{itemize}
For stem-wise activity control,
we run silence detection algorithm on the output audio
to get the activity sequence,
and compute the \textbf{frame-wise F1} score w.r.t.~the input control $\bm{a}_k$.
Such cycle consistency-based evaluation has also been used for existing temporal controls~\cite{wu2024music}.

% \slwu{First pass until here.}

\begin{table}[t]
    \centering
    \vspace{-6pt}
    \caption{
    Ablations on \textbf{core techniques}: \textit{stem grouping} and \textit{noise sharing}. 
    We generate all constituent stems in a mix in \textbf{one pass}.
    % at training and/or inference.
    % All constituent stems in a mix are inferenced in \textbf{one pass}.
    Our full setting, \textbf{C-(ii)}, generally outperforms all other ablated settings, with particularly strong gains on the more challenging MoisesDB dataset.
    ($n$ is the number of mixes in each evaluation dataset.)
    }
    \label{tab:main-abl}
\footnotesize
\renewcommand{\arraystretch}{1.2}
\setlength{\tabcolsep}{3.8pt}
\begin{tabular}{lccc|ccc}
\toprule
 & \multicolumn{2}{c}{\textbf{Train}} & \multicolumn{1}{c|}{\textbf{Infer.}} & \multicolumn{3}{c}{\textbf{MoisesDB} {\scriptsize ($n$ = 1488)} \textbf{/ {\color{CadetBlue}MusDB}} {\scriptsize {\color{CadetBlue}($n$ = 964)}}} \\
\cmidrule(lr){2-3} \cmidrule(lr){4-4} \cmidrule(lr){5-7}
    \textbf{Setting} \; & \shortstack{\textit{stem}\\[0.5ex]\textit{grp.}} & \shortstack{\textit{noise}\\\textit{share}} & \shortstack{\textit{noise}\\\textit{share}} & \shortstack{\textit{Stem Ctrl}\\[0.4ex]\textbf{FAD}$_\text{stem}$ $\downarrow$} & \shortstack{\textit{Mix Quality}\\[0ex]\textbf{FAD}$_\text{mix}$ $\downarrow$} & \shortstack{\textit{Mix Text Ctrl}\\[0.4ex]\textbf{CLAP} $\uparrow$} \\
\midrule
\textbf{A-(i)} & \xmark & \xmark & \xmark & 2.69 / {\color{CadetBlue}2.91} & 1.84 / {\color{CadetBlue}1.09} & 28.82 / {\color{CadetBlue}28.73} \\
\textbf{A-(ii)} & \xmark & \xmark & \cmark & 2.80 / {\color{CadetBlue}3.02} & 1.78 / {\color{CadetBlue}1.24} & 28.67 / {\color{CadetBlue}28.28} \\
\textbf{B-(i)} & \cmark & \xmark & \xmark & 2.41 / {\color{CadetBlue}2.92} & 1.55 / {\color{CadetBlue}\textbf{0.91}} & 28.85 / {\color{CadetBlue}29.14} \\
\textbf{B-(ii)} & \cmark & \xmark & \cmark & 2.41 / {\color{CadetBlue}2.97} & 1.53 / {\color{CadetBlue}1.10} & 28.93 / {\color{CadetBlue}28.76} \\ \cmidrule(l{0pt}r{0pt}){1-7}
\textbf{C-(ii)} & \cmark & \cmark & \cmark & \textbf{2.31} / {\color{CadetBlue}\textbf{2.72}} & \textbf{1.25} / {\color{CadetBlue}1.05} & \textbf{30.19} / {\color{CadetBlue}\textbf{29.27}} \\
\bottomrule
\end{tabular}
\end{table}

\begin{table}[t]
    \centering
\vspace{-6pt}
    \caption{
    \textbf{Workflow improvement} on generating $K$-stem mixes over a conventional iterative (\textbf{A-(i)}, $K$-passes) baseline.
    Our model, \textbf{C-(ii)}, offers the unique freedom to tradeoff speed for quality:
    the 1-pass setup provides a 50\%$+$ speedup, while the 2-pass setup achieves the overall best generations (\textbf{FAD}$_{\text{mix}}$ and \textbf{CLAP}) and still saves 25--50\% of total inference time.
    Stem groups \& prompts are from MoisesDB.
    % Number of mixes $n=$ 190 / 456 / 379 / 283 for $K=$ 3 / 4 / 5 / 6.
    % Results on \textit{workflow improvement} on MoisesDB evaluation set, faceted by the number of stems (i.e., $K$) in the mix.
    % The first row (\textbf{A-(i)}, $K$ passes) represents a possible workflow to generate cohesive mixes with existing models.
    % For the second row (\textbf{C-(ii)}, two passes), we generate half of the stems in each pass.
    % % ; in cases of odd $K$'s, the first pass takes one more stem.
    % Inference time (`\textbf{time} (s)') is the total runtime to output $K$ stems.
    % Overall, the two-pass workflow with our \textbf{C-(ii)} setting achieves the best generation quality, while saving 25--50\% of inference time compared to the conventional $K$-pass workflow.
    % full iterative vs.~1-pass vs.~2-pass. These are on MoisesDB examples, faceted over the number of stems (i.e., $K$) in the final mix.
    % Inference time (`\textbf{time} (s)') is the total runtime of all inference passes to get $K$ stems starting from scratch.
    }
    \label{tab:workflow}
    \footnotesize
    \setlength{\tabcolsep}{8.5pt}
    \renewcommand{\arraystretch}{1.2}
  
    \begin{tabular}{lc|ccc}
        \toprule
        & & \multicolumn{3}{c}{\Centerstack{$K = 3$ \, {\scriptsize ($n =$ 190)}  \, / \: {\color{BlueViolet}$K = 4$ \, {\scriptsize ($n =$ 456)}} \\{\color{Mahogany}$K = 5$ \, {\scriptsize ($n =$ 379)}} \: / \:\,{\color{OliveGreen}$K = 6$ \, {\scriptsize ($n =$ 283)}}}} \\
        \cmidrule(lr){3-5}
       \multirow{2}{*}{\textbf{Setting}} & \multirow{2}{*}{\Centerstack[l]{\# infer.\\passes}} & \multirow{2}{*}{\textbf{FAD}$_\text{mix}\downarrow$} & \multirow{2}{*}{\textbf{CLAP} $\uparrow$} & \multirow{2}{*}{\textbf{time} (s) $\downarrow$} \\
        & & & & \\
        \midrule
        \textbf{A-(i)}  & $K$ & \Centerstack{1.03 / {\color{BlueViolet}1.02}\\{\color{Mahogany}1.48} / {\color{OliveGreen}2.09}} & \Centerstack{\textbf{30.65} / {\color{BlueViolet}30.34}\\{\color{Mahogany}29.65} / {\color{OliveGreen}\textbf{30.67}}} & \Centerstack{4.10 / {\color{BlueViolet}5.50}\\{\color{Mahogany}6.88} / {\color{OliveGreen}8.28}} \\  \cmidrule(l{0pt}r{0pt}){1-5}
        \multirow{2}{*}[-1.5ex]{\textbf{C-(ii)}} & 2 & \Centerstack{\textbf{0.78} / {\color{BlueViolet}\textbf{0.83}}\\{\color{Mahogany}\textbf{1.34}} / {\color{OliveGreen}\textbf{1.92}}} & \Centerstack{30.62 / {\color{BlueViolet}\textbf{31.23}}\\{\color{Mahogany}\textbf{30.32}} / {\color{OliveGreen}30.05}} & \Centerstack{3.03 / {\color{BlueViolet}3.27}\\{\color{Mahogany}3.70} / {\color{OliveGreen}4.16}} \\\cmidrule(l{0pt}r{0pt}){2-5}
          & 1 & \Centerstack{1.06 / {\color{BlueViolet}1.12}\\{\color{Mahogany}1.56} / {\color{OliveGreen}2.29}} & \Centerstack{29.51 / {\color{BlueViolet}30.44}\\{\color{Mahogany}30.26} / {\color{OliveGreen}29.93}} & \Centerstack{\textbf{2.00} / {\color{BlueViolet}\textbf{2.52}} \\{\color{Mahogany}\textbf{3.13}} / {\color{OliveGreen}\textbf{3.60}}} \\[1ex]
        \bottomrule
    \end{tabular}
  \end{table}

\begin{table}[t]
    \centering
        \vspace{-6pt}
    \caption{
    Evaluating \textbf{stem activity control}.
    Our mechanism (Sec.~\ref{subsec:actcurve}) enables near-perfect activity control (framewise \textbf{F1}) with the flexibility to use it or not at inference, only at a moderate cost of other metrics.
    % Our mechanism (cf.~Sec.~?.?) enables a near-perfect activity control and the flexibility on whether or not to specify them at inference, only at a slight cost of other metrics.
    (\textbf{F1} is evaluated only on partially active stems.)
    % Results on stem activity control. 
    % Our conditioning mechanism achieves a near perfect activity control,
    % and grants the freedom on whether to specify them or not.
    % Moreover, this additional condition leads to only a slight degradation on other metrics.
    % (For activity control \textbf{Frame F1}, we only evaluate on cases where the input control is \textit{partially active}.)
    }
    \label{tab:actcurve}
    \scriptsize
    \renewcommand{\arraystretch}{1.2}
    \setlength{\tabcolsep}{4pt}
    \begin{tabular}{lcc|cccc}
        \toprule
        & \multicolumn{1}{c}{\textbf{Train}} & \multicolumn{1}{c|}{\textbf{Infer.}} & \multicolumn{4}{c}{\textbf{MoisesDB} {\tiny ($n$ = 1488)} \: / \: \textbf{ {\color{CadetBlue}MusDB}} {\tiny {\color{CadetBlue}($n$ = 964)}}} \\
        \cmidrule(lr){2-2} \cmidrule(l{1pt}r{1pt}){3-3} \cmidrule(lr){4-7}
        \textbf{Setting} & \shortstack{\textit{act}\\\textit{ctrl}} & \shortstack{\textit{act}\\\textit{ctrl}} & \shortstack{\textit{Stem Ctrl}\\[0.4ex]\textbf{FAD}$_\text{stem}$$\downarrow$} & \shortstack{\textit{Mix Quality}\\[0ex]\textbf{FAD}$_\text{mix}$$\downarrow$} & \shortstack{\textit{Mix Text Ctrl}\\[0.4ex]\textbf{CLAP}$\uparrow$} & \shortstack{\textit{Act Ctrl}\\[0.4ex]\textbf{F1} (\%)$\uparrow$} \\
        \midrule
        \textbf{C-(ii)} & \xmark & \xmark & \textbf{2.31} / {\color{CadetBlue}\textbf{2.72}} & \textbf{1.25} / {\color{CadetBlue}\textbf{1.05}} & \textbf{30.19} / {\color{CadetBlue}\textbf{29.27}} & {\color{LightGray}n.a.} \\ \hline
        \multirow{2}{*}{\Centerstack{\textbf{C-(ii)}\\$+$\textbf{Act}}} & \cmark & \xmark & 2.66 / {\color{CadetBlue}2.74} & 1.54 / {\color{CadetBlue}1.08} & 28.78 / {\color{CadetBlue}28.94} & {\color{LightGray}n.a.} \\
        & \cmark & \cmark & 2.47 / {\color{CadetBlue}2.77} & 1.46 / {\color{CadetBlue}1.13} & 29.55 / {\color{CadetBlue}29.14} & \textbf{99.42} / {\color{CadetBlue}\textbf{99.43}} \\
        \bottomrule
    \end{tabular}
\end{table}

%%%%%%%%%%%%%%%%%%%%%%%%%%%%%%%%%%%%%%%%%%%%%%%%%%%%%%%%%%%%%%%%%%%%%%%%%%%%%%%%
%%%%%%%%%%%%%%%%%%%%%%%%%%%%%%%%%%%%%%%%%%%%%%%%%%%%%%%%%%%%%%%%%%%%%%%%%%%%%%%%
%%%%%%%%%%%%%%%%%%%%%%%%%%%%%%%%%%%%%%%%%%%%%%%%%%%%%%%%%%%%%%%%%%%%%%%%%%%%%%%%
\section{Results \& Discussion}\label{sec:discuss}
Results of the core one-pass generation ablations are shown in Table~\ref{tab:main-abl}.
Our full \textbf{C-(ii)} setting overall produces the best generations,
especially on the more challenging MoisesDB evaluation set with more stems per mix.
Qualitatively, 
the baseline \textbf{A-(i)} setup consistently fails to produce synchronized stems, resulting in incoherent mixes.
The intermediate settings (\textbf{A-(ii), B-(i), B-(ii)}) show a marked improvement, occasionally producing musically-synchronized stems, despite being much less consistent than our full \textbf{C-(ii)} setup.
% Qualitatively,
% except for the \textbf{A-(i)} setup, whose output mixes sound mostly like combining unrelated stems together,
% all three other settings (i.e., \textbf{A-(ii), B-(i), B-(ii)}) can produce stems that are musically-synchronized to some extent, albeit less reliably than \textbf{C-(ii)}.
This gives some evidence that 
(\textit{1})~the initial noise does capture semantics in rhythm and harmony such that sharing noise only at inference (i.e, \textbf{*-(ii)}) enforces some degree of cohesiveness, and
(\textit{2})~the shared information among grouped stems present in setting \textbf{B-(*)}, e.g., text prompts,
also induces inter-stem cohesion, though being less effective than sharing the high-dimensional noise as in \textbf{C-(ii)}.

Table~\ref{tab:workflow} compares $K$-stem workflows between \textbf{A-(i)} (akin to individual-stem models~\cite{han2024instructme,nistal2024diff,chae2025mge}),
and those made possible by our \textbf{C-(ii)} setup.
The insights are: when given $K$ inference passes, \textbf{A-(i)}
can generate comparable or slightly better mixes than our `\textbf{C-(ii)}, 1-pass' workflow, but takes at least 2$\times$ as much time.
Meanwhile, our `\textbf{C-(ii)}, 2-pass' workflow achieves generally the best generations with consistently lower \textbf{FAD}$_\text{mix}$,
while still being 25--50\% faster than \textbf{A-(i)} with the time advantage increasing with stem count $K$.
We attribute the superior output quality of the 2-pass workflow to an inductive bias created at training---Since our subset sampling strategy (cf.~Sec.~\ref{subsec:data-group}) induces a binomial distribution of group sizes that peaks at $K/2$,
the model is inherently better at generating groups of this size.

Finally, Table~\ref{tab:actcurve} shows the performance of the model trained with stem-wise activity controls (cf.~Sec.~\ref{subsec:actcurve}).
The Frame F1 scores prove that our activity controls are near-perfect.
Other metrics take a slight hit, qualitatively due mainly to audio cleanliness degradation rather than
loss of inter-stem cohesion, as a tradeoff for the additional controls and the flexibility to optionally apply them at inference.
We encourage readers to visit our demo website {\footnotesize \url{https://stemphonic-demo.vercel.app}}, which showcases both static generations and \textit{composer-like} generations, where the authors act as a real-world user and iteratively shape the final mix.

% Straighforward tell the story from the tables.

% \noindent
% Some extra north stars:
% - explain why some ablations still work to some extent, e.g., in \textbf{B}, the text prompts provides some information on the groupings; also, whenever noise is bound \textbf{(ii)} at inference, the way the semantics captured in the latent space ensures some degree of alignment.

% \noindent
% - might have to hedge a bit for Table 3 (act curve control) FAD scores, since some of them drops to worse than ablations in Table 1. We can say that FAD packs lots of aspects in one score, stem alignment is one, audio cleanliness is also one, etc.

%%%%%%%%%%%%%%%%%%%%%%%%%%%%%%%%%%%%%%%%%%%%%%%%%%%%%%%%%%%%%%%%%%%%%%%%%%%%%%%%
%%%%%%%%%%%%%%%%%%%%%%%%%%%%%%%%%%%%%%%%%%%%%%%%%%%%%%%%%%%%%%%%%%%%%%%%%%%%%%%%
%%%%%%%%%%%%%%%%%%%%%%%%%%%%%%%%%%%%%%%%%%%%%%%%%%%%%%%%%%%%%%%%%%%%%%%%%%%%%%%%
\section{Conclusions \& Future Work}\label{sec:conclusion}
We proposed \textsc{Stemphonic},
a diffusion-/flow-based framework that enables one-pass generation of variable combinations of musically-synchronized stems.
% using a unified, stem-agnostic DiT backbone.
This was achieved with our proposed training-time interventions: stem grouping and noise sharing.
We further supported conditional multi-stem generation and stem-wise activity controls.
% combination of compatible stems in a single pass. 
Our experiments show that \textsc{Stemphonic} not only accelerates the multi-stem generation workflow by 25--50\% over the existing iterative baseline,  but also produces higher-quality mixes.

Our work opens several promising avenues for future research.
A deeper theoretical analysis of the noise sharing mechanism could yield fundamental insights into using high-dimensional noise as an explicit conditioning signal in generative models.
For user interaction, moving beyond simple stem type tags to free-text descriptions for individual stems remains an important next step for more precise control.
Furthermore, research into methods for controlling stem-wise musical novelty~\cite{bjare2024controlling}, or 
for agentically suggesting stem combination presets~\cite{deng2024composerx} given a global text prompt,
can significantly enhance user engagement and collaborative potential of such systems.

% \slwu{First pass done!!}
% Future work ideas:\\
% - Theories around noise sharing

% \noindent
% - More descriptive stem control (i.e., stem type tags $\rightarrow$ free text)

% \noindent
% - Novelty control with some text or symbolic descriptors

% \noindent
% - Stem combination preset generation for less initial friction

% \clearpage
\bibliographystyle{IEEEbib}
\footnotesize
\bibliography{strings,refs}

@inproceedings{rombach2022high,
  title={High-resolution image synthesis with latent diffusion models},
  author={Rombach, Robin and Blattmann, Andreas and Lorenz, Dominik and Esser, Patrick and Ommer, Bj{\"o}rn},
  booktitle={CVPR},
  year={2022}
}

@article{song2020denoising,
  title={Denoising diffusion implicit models},
  author={Song, Jiaming and Meng, Chenlin and Ermon, Stefano},
  journal={arXiv:2010.02502},
  year={2020}
}

@inproceedings{kim2025amuse,
  title={Amuse: Human-{AI} Collaborative Songwriting with Multimodal Inspirations},
  author={Kim, Yewon and Lee, Sung-Ju and Donahue, Chris},
  booktitle={CHI},
  year={2025}
}

@inproceedings{donahue2025hookpad,
  title={Hookpad {Aria}: A Copilot for Songwriters},
  author={Donahue, Chris and Wu, Shih-Lun and Kim, Yewon and Carlton, Dave and Miyakawa, Ryan and Thickstun, John},
  booktitle={ISMIR Late-breaking Demos},
  year={2024}
}

@article{wu2024music,
  title={{Music ControlNet}: Multiple time-varying controls for music generation},
  author={Wu, Shih-Lun and Donahue, Chris and Watanabe, Shinji and Bryan, Nicholas J.},
  journal={IEEE/ACM T-ASLP},
  year={2024}
}

@inproceedings{tsai2025musecontrollite,
  title={{MuseControlLite}: Multifunctional Music Generation with Lightweight Conditioners},
  author={Tsai, Fang-Duo and Wu, Shih-Lun and Lee, Weijaw and Yang, Sheng-Ping and Chen, Bo-Rui and Cheng, Hao-Chung and Yang, Yi-Hsuan},
  booktitle={ICML},
  year={2025}
}

@inproceedings{copet2023simple,
  title={Simple and controllable music generation},
  author={Copet, Jade and Kreuk, Felix and Gat, Itai and Remez, Tal and Kant, David and Synnaeve, Gabriel and Adi, Yossi and D{\'e}fossez, Alexandre},
  booktitle={NeurIPS},
  year={2023}
}

@inproceedings{evans2024fast,
  title={Fast timing-conditioned latent audio diffusion},
  author={Evans, Zach and Carr, CJ and Taylor, Josiah and Hawley, Scott H and Pons, Jordi},
  booktitle={ICML},
  year={2024}
}

@article{xu2025qwen2,
  title={Qwen2.5-omni technical report},
  author={Xu, Jin and Guo, Zhifang and He, Jinzheng and Hu, Hangrui and He, Ting and Bai, Shuai and Chen, Keqin and Wang, Jialin and Fan, Yang and Dang, Kai and others},
  journal={arXiv:2503.20215},
  year={2025}
}

@inproceedings{kilgour2019fr,
  title={Fr\'echet audio distance: A metric for evaluating music enhancement algorithms},
  author={Kilgour, Kevin and Zuluaga, Mauricio and Roblek, Dominik and Sharifi, Matthew},
  booktitle={Interspeech},
  year={2019}
}

@inproceedings{hershey2017cnn,
  title={{CNN} architectures for large-scale audio classification},
  author={Hershey, Shawn and Chaudhuri, Sourish and Ellis, Daniel PW and Gemmeke, Jort F and Jansen, Aren and others},
  booktitle={ICASSP},
  year={2017}
}

@inproceedings{casebeer2025,
  title={A Generative-First Neural Audio Autoencoder},
  author={Casebeer, Jonah and Zhu, Ge and Wang, Zhepei and Bryan, Nicholas J.},
  booktitle={ICASSP},
  year={2026}
}

@inproceedings{wu2023large,
  title={Large-scale contrastive language-audio pretraining with feature fusion and keyword-to-caption augmentation},
  author={Wu, Yusong and Chen, Ke and Zhang, Tianyu and Hui, Yuchen and Berg-Kirkpatrick, Taylor and Dubnov, Shlomo},
  booktitle={ICASSP},
  year={2023}
}

@article{loshchilov2017decoupled,
  title={Decoupled weight decay regularization},
  author={Loshchilov, Ilya and Hutter, Frank},
  journal={arXiv:1711.05101},
  year={2017}
}

@article{raffel2020exploring,
  title={Exploring the limits of transfer learning with a unified text-to-text transformer},
  author={Raffel, Colin and Shazeer, Noam and Roberts, Adam and Lee, Katherine and Narang, Sharan and Matena, Michael and Zhou, Yanqi and Li, Wei and Liu, Peter J},
  journal={Journal of Machine Learning Research (JMLR)},
  year={2020}
}

@article{novack2024ditto,
  title={{DITTO}: Diffusion inference-time t-optimization for music generation},
  author={Novack, Zachary and McAuley, Julian and Berg-Kirkpatrick, Taylor and Bryan, Nicholas J.},
  journal={ICML},
  year={2024}
}

@inproceedings{tal2024joint,
  title={Joint audio and symbolic conditioning for temporally controlled text-to-music generation},
  author={Tal, Or and Ziv, Alon and Gat, Itai and Kreuk, Felix and Adi, Yossi},
  booktitle={ISMIR},
  year={2024}
}

@inproceedings{tsai2024audio,
  title={Audio {P}rompt {A}dapter: Unleashing music editing abilities for text-to-music with lightweight finetuning},
  author={Tsai, Fang-Duo and Wu, Shih-Lun and Kim, Haven and Chen, Bo-Yu and Cheng, Hao-Chung and Yang, Yi-Hsuan},
  booktitle={ISMIR},
  year={2024}
}

@article{team2025live,
  title={Live Music Models},
  author={Caillon, Antoine and McWilliams, Brian and Tarakajian, Cassie and Simon, Ian and Manco, Ilaria and Engel, Jesse and Constant, Noah and Li, Pen and Denk, Timo I and others},
  journal={arXiv:2508.04651},
  year={2025}
}

@inproceedings{novack2025presto,
  title={Presto! Distilling Steps and Layers for Accelerating Music Generation},
  author={Novack, Zachary and Zhu, Ge and Casebeer, Jonah and McAuley, Julian and Berg-Kirkpatrick, Taylor and Bryan, Nicholas J.},
  booktitle={ICLR},
  year={2025}
}

@article{agostinelli2023musiclm,
  title={{MusicLM}: Generating music from text},
  author={Agostinelli, Andrea and Denk, Timo I and Borsos, Zal{\'a}n and Engel, Jesse and others},
  journal={arXiv:2301.11325},
  year={2023}
}

@inproceedings{malandro2023composer,
  title={Composer's {A}ssistant: An Interactive Transformer for Multi-Track MIDI Infilling},
  author={Malandro, Martin E},
  booktitle={ISMIR},
  year={2023}
}

@inproceedings{malandro2024composer,
  title={Composer's {A}ssistant 2: Interactive Multi-Track MIDI Infilling with Fine-Grained User Control},
  author={Malandro, Martin E},
  booktitle={ISMIR},
  year={2024}
}

@inproceedings{bock2016madmom,
  title={Madmom: A new python audio and music signal processing library},
  author={B{\"o}ck, Sebastian and Korzeniowski, Filip and Schl{\"u}ter, Jan and Krebs, Florian and Widmer, Gerhard},
  booktitle={ACM MM},
  year={2016}
}

@inproceedings{evans2025stable,
  title={Stable {A}udio {O}pen},
  author={Evans, Zach and Parker, Julian D and Carr, CJ and Zukowski, Zack and Taylor, Josiah and Pons, Jordi},
  booktitle={ICASSP},
  year={2025}
}

@article{ho2022classifier,
  title={Classifier-free diffusion guidance},
  author={Ho, Jonathan and Salimans, Tim},
  journal={arXiv:2207.12598},
  year={2022}
}

@inproceedings{mariani2024multi,
  title={Multi-Source Diffusion Models for Simultaneous Music Generation and Separation},
  author={Mariani, Giorgio and Tallini, Irene and Postolache, Emilian and Mancusi, Michele and Cosmo, Luca and Rodol{\`a}, Emanuele},
  booktitle={ICLR},
  year={2024}
}

@inproceedings{han2024instructme,
  title={{InstructME}: An instruction guided music edit and remix framework with latent diffusion models},
  author={Han, Bing and Dai, Junyu and Hao, Weituo and He, Xinyan and Guo, Dong and Chen, Jitong and Wang, Yuxuan and Qian, Yanmin and Song, Xuchen},
  booktitle={IJCAI},
  year={2024}
}

@inproceedings{yao2025jen,
  title={{JEN-1} {C}omposer: A unified framework for high-fidelity multi-track music generation},
  author={Yao, Yao and Li, Peike and Chen, Boyu and Wang, Alex},
  booktitle={AAAI},
  year={2025}
}

@inproceedings{parker2024stemgen,
  title={{StemGen}: A music generation model that listens},
  author={Parker, Julian and Spijkervet, Janne and Kosta, Katerina and Yesiler, Furkan and Kuznetsov, Boris and Wang, Ju-Chiang and Avent, Matt and Chen, Jitong and Le, Duc},
  booktitle={ICASSP},
  year={2024}
}

@inproceedings{nistal2024diff,
  title={{Diff-A-Riff}: Musical accompaniment co-creation via latent diffusion models},
  author={Nistal, Javier and Pasini, Marco and Aouameur, Cyran and Grachten, Maarten and Lattner, Stefan},
  booktitle={ISMIR},
  year={2024}
}

@inproceedings{karchkhadze2025simultaneous,
  title={Simultaneous music separation and generation using multi-track latent diffusion models},
  author={Karchkhadze, Tornike and Izadi, Mohammad Rasool and Dubnov, Shlomo},
  booktitle={ICASSP},
  year={2025}
}

@inproceedings{rouard2025musicgen,
  title={{MusicGen-Stem}: Multi-stem music generation and edition through autoregressive modeling},
  author={Rouard, Simon and San Roman, Robin and Adi, Yossi and Roebel, Axel},
  booktitle={ICASSP},
  year={2025}
}

@inproceedings{strano2025stage,
  title={STAGE: Stemmed Accompaniment Generation through Prefix-Based Conditioning},
  author={Strano, Giorgio and Ballanti, Chiara and Crisostomi, Donato and Mancusi, Michele and Cosmo, Luca and Rodol{\`a}, Emanuele},
  booktitle={ISMIR},
  year={2025}
}

@inproceedings{chae2025mge,
  title={{MGE-LDM}: Joint Latent Diffusion for Simultaneous Music Generation and Source Extraction},
  author={Chae, Yunkee and Lee, Kyogu},
  booktitle={NeurIPS},
  year={2025}
}

@inproceedings{peebles2023scalable,
  title={Scalable diffusion models with {T}ransformers},
  author={Peebles, William and Xie, Saining},
  booktitle={CVPR},
  year={2023}
}

@inproceedings{esser2024scaling,
  title={Scaling rectified flow transformers for high-resolution image synthesis},
  author={Esser, Patrick and Kulal, Sumith and Blattmann, Andreas and Entezari, Rahim and M{\"u}ller, Jonas and others},
  booktitle={ICML},
  year={2024}
}

@inproceedings{liu2023flow,
  title={Flow Straight and Fast: Learning to Generate and Transfer Data with Rectified Flow},
  author={Liu, Xingchao and Gong, Chengyue and others},
  booktitle={ICLR},
  year={2023}
}

@inproceedings{kynkaanniemi2024applying,
  title={Applying guidance in a limited interval improves sample and distribution quality in diffusion models},
  author={Kynk{\"a}{\"a}nniemi, Tuomas and Aittala, Miika and Karras, Tero and Laine, Samuli and Aila, Timo and Lehtinen, Jaakko},
  booktitle={NeurIPS},
  year={2024}
}

@misc{musdb18,
  author       = {Rafii, Zafar and
                  Liutkus, Antoine and
                  Fabian-Robert St{\"o}ter and
                  Mimilakis, Stylianos Ioannis and
                  Bittner, Rachel},
  title        = {The {MUSDB18} corpus for music separation},
  year         = 2017,
  url          = {https://doi.org/10.5281/zenodo.1117372}
}

@inproceedings{pereira2023moisesdb,
  title={{MoisesDB}: A dataset for source separation beyond 4-stems},
  author={Pereira, Igor and Ara{\'u}jo, Felipe and Korzeniowski, Filip and Vogl, Richard},
  booktitle={ISMIR},
  year={2023}
}

@inproceedings{ciranni2025cocola,
  title={{COCOLA}: Coherence-oriented contrastive learning of musical audio representations},
  author={Ciranni, Ruben and Mariani, Giorgio and Mancusi, Michele and Postolache, Emilian and Fabbro, Giorgio and Rodol{\`a}, Emanuele and Cosmo, Luca},
  booktitle={ICASSP},
  year={2025}
}

@inproceedings{bjare2024controlling,
  title={Controlling surprisal in music generation via information content curve matching},
  author={Bjare, Mathias Rose and Lattner, Stefan and Widmer, Gerhard},
  booktitle={ISMIR},
  year={2024}
}

@article{deng2024composerx,
  title={Composer{X}: Multi-agent symbolic music composition with {LLM}s},
  author={Deng, Qixin and Yang, Qikai and Yuan, Ruibin and others},
  journal={arXiv:2404.18081},
  year={2024}
}

\end{document}